# Quantum Mechanics and Motion:
# A Modern Perspective

Gerald E. Marsh

Argonne National Laboratory (Ret)
5433 East View Park
Chicago, IL 60615

E-mail: gemarsh@uchicago.edu

**Abstract.** This essay is an attempted to address, from a modern perspective, the motion of a particle. Quantum mechanically, motion consists of a series of localizations due to repeated interactions that, taken close to the limit of the continuum, yields a world-line. If a force acts on the particle, its probability distribution is accordingly modified. This must also be true for macroscopic objects, although now the description is far more complicated by the structure of matter and associated surface physics.

The elements that comprise this essay are based on well-founded and accepted physical principles—but the way they are put together, as well as the view of commonly accepted forces and the resulting motion of macroscopic objects that emerges, is unusual. What will be shown is that classical motion can be identified with collective quantum mechanical motion. Not very surprising, but the conception of motion that emerges is somewhat counterintuitive. After all, we all know that the term $\hbar^2/2m$ in the Schrödinger equation becomes ridiculously small for *m* corresponding to a macroscopic object.

**Space-time and Quantum Mechanics**

To deal with the concept of motion we must begin with the well-known problem of the inconsistency inherent in the melding of quantum mechanics and special relativity. One of the principal examples that can illustrate this incompatibility is the Minkowski diagram, where well-defined world-lines are used to represent the paths of elementary particles while quantum mechanics disallows the existence of any such well defined world-lines. Despite this conceptual dissonance, the fusion of quantum mechanics and special relativity has proved to be enormously fruitful. This point has been made by Sklar[1] in his book *Space, Time, and Spacetime*: "Despite the rejection in quantum theory of the very notions used in the original justification of the construction of the space-time of special relativity, it is still possible to formulate quantum theory in terms of the space-time constructed in special relativity."

Feynman[2] in his famous paper "The Theory of Positrons" partially avoids the above conundrum, implicit in drawing space-time diagrams, by observing that solutions to the Schrödinger and Dirac equations can be visualized as describing the scattering of a plane wave by a potential. In the case of the Dirac equation, the scattered waves may proceed both forward and backward in time and may suffer further scattering by the same or other potentials. An identity is made between the negative energy components of the scattered wave and the waves traveling backward in time. This interpretation is valid for both virtual and real particles. While one generally does not indicate the waves, and instead draws world-lines in Minkowski space between such scatterings, it is generally understood that the particle represented by these waves does not have a well defined location in space or time between scatterings.[3]

The Feynman approach visualizes a non-localized plane wave impinging on a region of space-time containing a potential, and the particle the wave represents being localized[4] to a finite region of Minkowski space by interaction with the potential. The waves representing the scattered particle subsequently spread through space and time until there is another interaction in the same potential region or in a different region also containing a potential, again localizing the particle. Even this picture is problematic since the waves are not observable between interactions. For the Dirac equation, the now famous Figure 1 is intended to represent electron scattering from two different regions containing a scattering potential. The plane electron wave comes in from the lower left of the figure, is scattered by the potential at A(3). (a) shows the scattered wave going both forward and backward in time; (b) and (c) show two second order processes where (b) shows a normal scattering forward in time and (c) the possibility of pair production. Feynman meant this figure to apply to a virtual process, but—as discussed by Feynman—with the appropriate interpretation it applies to real pair production as well. Although the lines are drawn to represent these particles, no well-defined world-lines exists.

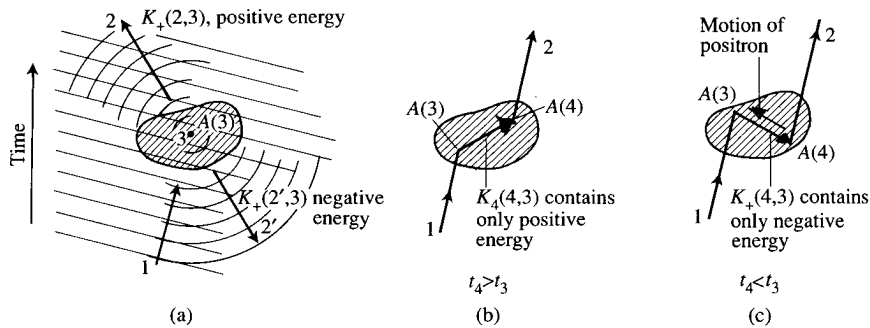

Figure 1. Different electron scattering possibilities from a potential region. (a) is a first order process while (b) and (c) are second order. [Based on Figure 2 of R. P. Feynman, "Theory of Positrons", *Phys. Rev.* **76**, 749-759 (1949)]

In a bubble chamber, where the path followed by the charged particles is made visible by repeated localizing interactions with the medium, one would observe a pair creation event at A(4), an electron coming in from the lower left, and an annihilation event at A(3). Of course, since the particles involved here are massive, in the case of real pair production the interval between A(3) and A(4) is time-like and the spatial distance between these events depends on the observer.

To reiterate, a world-line is a classical concept that is only approximated in quantum mechanics by the kind of repeated interactions that make a path visible in a bubble chamber.[5] Minkowski space is the space of *events*—drawing a world-line in a Minkowski diagram implicitly assumes such repeated interactions taken to the limit of the continuum.[6] While the characterization of Minkowski space as the space of events is often obscured by drawing world-lines as representing the putative path of a particle in space-time independent of its interactions, remembering that each point in Minkowski space is the position of a potential event removes much of the apparent incompatibility between quantum mechanics and special relativity, but it leaves us with a revised view of what constitutes motion.

**Quantum Mechanical Motion**

The picture of motion that emerges after the melding of quantum mechanics and special relativity is very unlike that of the classical picture of the path of a massive particle—like a marble—moving in space-time. Consider a Minkowski diagram showing the world-lines of several marbles at different locations. Given a space-like hypersurface corresponding to an instant of time in some frame, all the marbles would be visible at some set of locations. If one chooses a neighboring instant of time, these marbles would all still be visible at slightly different locations. This is because of the sharp localization of the marbles in space and time due to the continual interactions of their constituent components. Now consider the case of several elementary particles such as electrons. On any space-like hypersurface, the only particles "visible" would be those that were localized by an interaction to a region of space-time that included the instant of time corresponding to the hypersurface.[7] After any localization, the wave function of a particle spreads both in space and in either direction in time. Consequently, neighboring hypersurfaces (in the same reference frame) corresponding to slightly different times could have a different set of particles that were "visible." If motion consists of a sequential series of localizations along a particle's path, it is not possible to define a continuum of movement in the classical sense—there exists only a series of "snapshots."

Haag,[8] has put this somewhat different terms: "The resulting ontological picture differs drastically from a classical one. It sketches a world, which is continuously evolving, where new facts are permanently emerging. Facts of the past determine only probabilities of future possibilities. While an individual event is considered as a real fact, the correlations between events due to quantum mechanical entanglement imply that an individual object can be regarded as real only insofar as it carries a causal link between two events. The object remains an element of potentiality as long as the target result has not become a completed fact."

It is important to emphasize that between localizations due to interactions, an elementary particle does not have a specifiable location, although—because it is located with very high probability[9] somewhere within the future and past light cones associated with its most recent localization—it would contribute to the local mass-energy density. This is not a matter of our ignorance, it is a fundamental property of quantum mechanics; Bell's theorem tells us that there are no hidden variables that could specify a particle's position between localizations.

As an example of how localization works, consider a single atom. Its nucleus is localized by the continuous interactions of its constituent components. The electrons are localized due to interactions with the nucleus, but only up to the appropriate quantum numbers—$n$, $l$, $m$, and $s$. One cannot localize the electrons to positions in their "orbits."

Implicit in the discussion above is that an "elementary particle" is not a "particle" in the sense of classical physics. The advent of quantum mechanics mandated that the classical notion of a particle be given up. But rather than accept this, there were many attempts in the 20th century to retain the idea of a classical particle by a mix of classical and quantum mechanical concepts. Perhaps the best was David Bohm's 1952 theory that introduced the idea of a "quantum potential". None of these were really successful. In the end, we must live with the fact that an elementary particle is some form of space-time excitation that can be localized through interactions and even when not localized obeys all the relevant conservation rules and retains the "particle" properties like mass, spin, charge, etc.

Above, the flat space-time of special relativity was used in the discussion. When the space-time curvature due to gravitation is included, Minkowski diagrams become almost impossible to draw: Given a space-like hypersurface, the rate of clocks at any point on the hypersurface depends on the local mass-energy density and on local charge. Compared to a clock in empty space-time, a clock near a concentration of mass-energy will run slower and will run faster near an electric charge of either sign. Thus the hypersurface does not remain "planar" as it evolves in time. To draw world-lines one must take into account the general relativistic metric. This is why one uses light cone indicators at points contained in regions of interest.

The concepts of quantum mechanical localization and the resulting picture of motion are especially important in discussing many-particle problems and the transition to the classical world. In considering the penetration of a potential barrier, for example, one often restricts the problem to a single particle and calculates the probability that it will be found on the far side of the barrier. For the many-particle case, say the surface barrier of a metal treated as a free-electron gas in a smeared positive background—an example that will be relevant later in this essay—one would find that those electron wave functions that have been localized on the far side of the barrier will contribute to a real negative charge density. This charge density will interact with the smeared positive background.

**Force, Fields, and Motion**

Fields in classical physics are defined in terms of forces on either massive particles—in the case of Newtonian mechanics, or charges in the case of electromagnetism. General Relativity changed our way of thinking about the gravitational field by replacing the concept of a force field with the curvature of space-time.

Starting with Einstein and Weyl,[10] there have been many attempts to geometrize electromagnetic forces. In all these attempts, charge—like mass in Newtonian mechanics—is treated as an irreducible element of electromagnetic theory that must be introduced *ab initio*. Its origin is not properly a part of the theory. It does, nevertheless, have a unique space-time signature.[11] Charge of either sign causes a *negative* curvature of space-time. The Einstein-Maxwell system of equations does not, however, allow different *geometric* representations for the electric fields due to positive and negative charges. This is a direct result of the fact that the sources of the Einstein-Maxwell system are embodied in the energy-momentum tensor, which depends only on the (non-gravitational) energy density. Charge, in its geometrical effect on space-time, always enters as $Q^2$ so that both positive and negative charges affect space-time in the same way. Consequently, the electric field due to positive and negative charges cannot be identified with distinct changes in space-time geometry. This also follows from the fact that, if we ignore the very small curvature of space-time due to the energy density of the field, only charged particles are directly affected by the presence of an electromagnetic field. Thus, a full geometrization of charge does not appear to be possible within the framework of the Einstein-Maxwell equations.

The advent of modern gauge theory, incorporating the concepts of symmetry breaking and compensation fields, radically changed the understanding of fields. The electromagnetic interaction of charged particles in particular could be interpreted in terms of a local—as opposed to global—gauge theory within the framework of quantum mechanics. Interpreting the electromagnetic field as a local gauge field takes into account the existence of positive and negative charges and gives a good representation of the electromagnetic forces. It also gives us a concept of the electric field somewhat more enlightening than the classical one where the field is defined as the ratio of the force on test charge to the charge in the limit that the charge goes to zero.

The key concept for representing the electromagnetic force as a gauge field is the recognition that the phase of a particle's wave function must be treated as a new physical degree of freedom dependent on the particle's space-time position. The 4-dimensional vector potential plays the role of a connection relating the phase from point-to-point. Thus, the vector potential becomes the fundamental field for electromagnetism. The Aharonov and Bohm effect is generally cited to prove that this potential can produce observable effects, thereby confirming its reality.

The "gauge principle", as it is often called, is well illustrated by considering the non-relativistic Schrödinger equation in the context of electromagnetism.[12] It is also possible to also give a relativistic version of the argument that appears below.

The Schrödinger equation for a free particle,

$$-\frac{\hbar^2}{2m}\nabla^2 \Psi(\vec{x},t) = i\hbar\, \partial_t \Psi(\vec{x},t),$$

is not invariant under the *local* phase transformation

$$\Psi(\vec{x},t) \rightarrow \Psi'(\vec{x},t) = e^{i\alpha(\vec{x},t)}\Psi(\vec{x},t).$$

To be invariant under such a transformation, the free particle Schrödinger equation must be modified so that it no longer represents a free particle, but rather one moving under the influence of a force. For the case of electromagnetism, the free particle Schrödinger equation must be replaced by

$$\left[\frac{1}{2m}\left(-i\hbar\nabla - q\vec{A}\right)^2 + qV\right]\Psi(\vec{x},t) = i\hbar\,\partial_t\Psi(\vec{x},t),$$

where $\vec{A}$ and V transform according to

$$\vec{A} \rightarrow \vec{A}' = \vec{A} + q^{-1}\nabla\alpha(\vec{x},t)$$

$$V \rightarrow V' = V - q^{-1}\partial_t\alpha(\vec{x},t)$$

when $\Psi(\vec{x}, t) \to \Psi'(\vec{x}, t)$.

The essence of the "gauge principle" is that demanding invariance under a local phase transformation corresponds to the introduction of a force. Of course, one can argue in the reverse: the introduction of a force can be represented as a local phase transformation. A simple example will be given below.

A free particle at rest samples a volume of space *at least* as large as its Compton wavelength, and the wave function associated with this sampling is such that a spherical volume is sampled in the absence of external forces. One might think here of a Gaussian packet (the lowest order wave function for the simple harmonic oscillator) which has the property of minimizing the uncertainty in both *x* and *p* thereby giving the maximum localization possible.

If a force acts on the particle—say along the *x*-axis—this symmetry is broken by an extension of the probability distribution (the volume sampled) along the *x*-axis. To actually be "seen" to move, the particle must participate in a series of interactions so as to repeatedly localize it along its path of motion. If the force acting on the particle is modeled as a virtual exchange of quanta, such an exchange—viewed as an interaction—would serve to localize the particle. The propagation of a Gaussian wave packet representing the propagation of a charged particle under the influence of a constant force is an example well worth discussing further. This problem has recently been extensively treated by Robinett[13] and Vandegrift.[14]

The Gaussian wave packet $\psi_0(x,t)$ is a solution to the free-particle, one-dimensional Schrödinger equation

$$i\hbar \frac{\partial \Psi}{\partial t} = -\frac{\hbar^2}{2m} \frac{\partial^2 \Psi}{\partial x^2} - Fx\Psi$$

with $F = 0$. This solution has the property that it will remain centered at $x = 0$ for all values of $t$. Now let $F$ be a time-independent, uniform force, implying a constant acceleration. In classical mechanics such a force has the kinematic relation

$$x(t) = x_0 + v_0 t + \frac{1}{2}at^2,$$

where $x_0$ and $v_0$ are the initial position and velocity, and $a$ is the acceleration. What Vandegrift shows is that the Gaussian packet solution to the Schrödinger equation with $F$ a uniform force becomes a wave packet centered at $x(t)$, that is

$$\Psi(x,t) = \psi_0\left(x - x_0 - v_0 t - \frac{1}{2}at^2, t\right) e^{iS(x,t)}$$

is a solution to the Schrödinger equation. The phase $e^{iS(x,t)}$ is a local phase transformation corresponding to $e^{i\alpha(x,t)}$ above, and $S(x,t)$ is explicitly given by

$$\frac{\hbar}{m} S(x,t) = v_0 x + axt - \frac{1}{2}av_0 t^2 - \frac{1}{6}a^2 t^3 - \frac{1}{2}v_0^2 t.$$

This solution to Schrödinger's equation shows that the imposition of a uniform force is equivalent to making a non-relativistic transformation to an accelerating reference frame. It is also an example of the gauge principle.

**Quantum Electrostatics**

The gauge principle should also be able to explain macroscopic phenomena. The example to be used here will be that of electrostatics. Discussing electrostatics in a quantum mechanical framework is perhaps one of the most counterintuitive examples of collective quantum phenomena leading to classical behavior. What will be shown here is that the electric field, best interpreted as a phase field, affects the electron wave functions at the surface of a conductor and collectively this is what is responsible for the force acting on the conductor. Of course one could simply use the classical electric field concept to achieve the same result, but—recalling the example of the Gaussian packet— greater insight into how the classical motion emerges is gained by thinking of the electric field as a phase field.

If a sphere holding a net positive charge $Q$ is placed in an initially uniform electric field $E_0$, it will experience a force in the direction of the applied field. In solving Laplace's equation in terms of spherical harmonics, this force results from the term $Q/4\pi a^2$, where $a$ is the radius of the sphere. The total charge density on the surface of the sphere is $3\varepsilon_0 E_0 \cos\theta + Q/4\pi a^2$. The electric field lines and associated surfaces of constant potential are shown in Fig. 2.

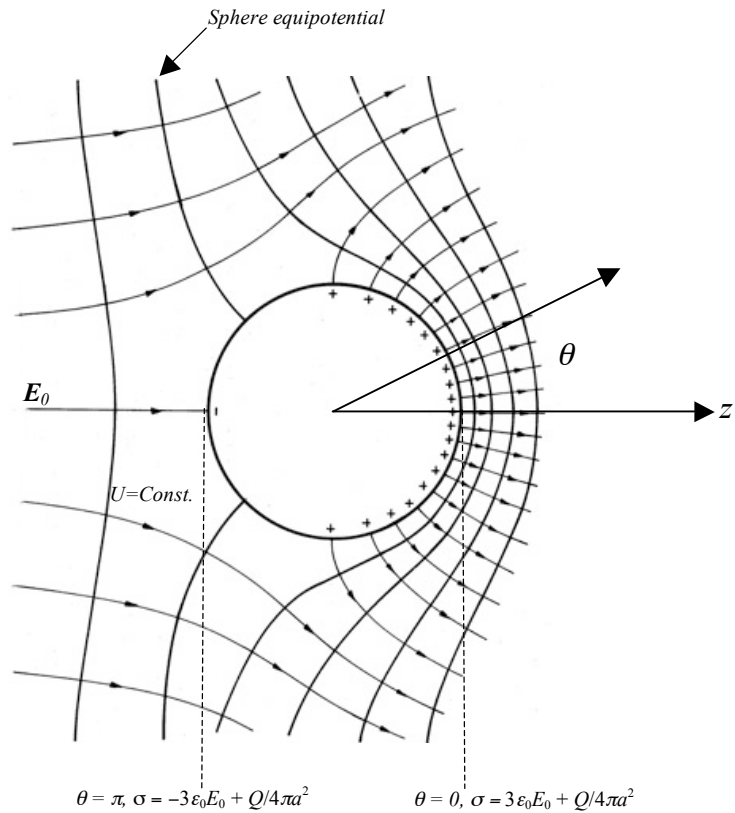

Figure 2. Electric field $E_0$ and associated surfaces of constant potential $U$ for a positively charged sphere of radius $a$ in an initially uniform field. The induced surface charge density varies with $\theta$, whereas that due to $Q$ does not.

Notice that the constant potential surface corresponding to the potential of the sphere intersects the sphere and divides its surface so that those electric field lines terminating on negative surface charges are on one side of the intersection, and those whose origin is on positive surface charges are on the other. The value of $\theta$ giving the location of the intersection is given by the solution to the equation

$$\cos\theta + \frac{Q}{12\pi\varepsilon_0 E_0 a^2} = 0.$$

For $Q > 12\pi\varepsilon_0 E_0 a^2$, the electric field is directed outward from the surface for all $\theta$. As is readily seen from the figure, $\nabla U\big|_{\theta=0} > \nabla U\big|_{\theta=\pi}$, which implies a net force in the positive z-direction.

From a quantum mechanical perspective, the wave function at the surface is modified by the electric field interpreted as a phase field—similar to the example of the Gaussian wave packet discussed above. A net positive charge corresponds to removing a portion of the electron cloud of the nuclei near the surface, thereby unshielding these nuclei, which are the source of the positive charge. The wave function at the surface, as will be seen, is affected asymmetrically by the presence of an external electric field.

In order to calculate the wave function one has to simplify the problem and often uses the so-called jellium model[15] where the metal is modeled as a uniform positive background and an interacting electron gas. The surface of the metal is represented by the jellium (or geometrical) edge and is located at one half of the lattice spacing from the surface atom nuclei. The rapidly decaying electron cloud density extends beyond the geometrical surface.

The centroid of the excess charge distribution[†] (also known as the electrical surface) linearly induced by an external electric field is given by

$$z_{ref} = \frac{\int_{-\infty}^{\infty} x\, n_\sigma(x)\, dx}{\int_{-\infty}^{\infty} n_\sigma(x)\, dx}.$$

---

[†] The excess charge distribution is also known as the screening or induced charge distribution.

Here $n_\sigma(x)$ is the surface-charge density induced by the electric field perpendicular to the surface.[16] This centroid, calculated for $0 \le \theta \le 2\pi$, gives the position of the electrical surface—that surface where the external electric field appears to start. This surface is also the analog of the image plane (for the jellium and excess charge distribution) in the case of a plane conductor. Because we will be considering the position of the electrical surface at $\theta = 0$ and $\theta = \pi$, its position will be denoted by $z_{ref}$. The geometry and notional wave functions are shown in Fig. 3.

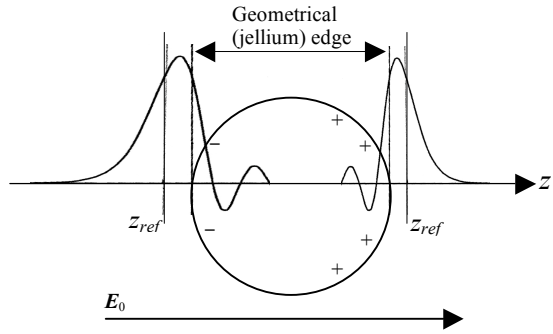

Fig. 3. Surface wave functions for a positively charged metal sphere in an *initially* uniform electric field $E_0$. $z_{ref}$ is the centroid of the excess charge distribution at $\theta = 0$ and $\theta = \pi$. Notice that the location of $z_{ref}$ is closer to the geometrical edge on positively charged portion of the sphere. The actual magnitude of the distance $z_{ref}$ is around 3 a.u., or about 1.6 Å.

The symmetry of the electron-probability distribution along the x-axis is broken by the charge $Q$. This results in a change in the position of $z_{ref}$, which is determined by the net electric field due to the charge on the sphere and the external electric field.

Thus, for an uncharged sphere placed in an initially uniform field, classically the gradient of the potential giving the field near the surface at $\theta = 0$ and $\theta = \pi$ is equal in magnitude, but opposite in direction with respect to the surface of the sphere. The net force therefore vanishes. Quantum mechanically, the location of the image surface is at the same distance from the jellium edge at $\theta = 0$ and $\theta = \pi$ so that the excess charge distribution interacts with the jellium equally yielding no net force in the direction of the field.

This will no longer be the case if the sphere is charged: the electric field at the surface due to the charge will asymmetrically sum with that due to the external field $\boldsymbol{E}_0$. For a positively charged sphere the net field at $\theta = 0$ will be greater than at $\theta = \pi$. As the magnitude of the external electric field increases, the image surface moves inward towards the surface[17] at $\theta = 0$, but less so than at $\theta = \pi$. Because the electrical image surface is now closer to the jellium at $\theta = 0$ than at $\theta = \pi$, there is a net force in the positive $z$-direction (the positive charge on the sphere can be pictured as residing on the geometrical surface).

For a negatively charged sphere, the image surface moves outward (away from the surface) as the magnitude of the external field increases,[18] but more so at $\theta = \pi$ than at $\theta = 0$. The negative charge on the geometrical surface is then further away from the effective negative charge due to the excess charge distribution. The separation is greater at $\theta = \pi$ than at $\theta = 0$. This results in a net force in the negative $z$-direction.

Quantum mechanically, the origin of the force is similar to the example given earlier of the Gaussian packet, but in the case of the more complicated problem of a charged macroscopic sphere, one must adopt some simplifying model of the surface and its associated wave function. Above, the jellium model was used for the surface of a charged sphere with notional electron wave functions to illustrate the origin of the classical force. Thus, the collective force due to the asymmetric excess charge distribution that results from the localization of the underlying electron wave functions *is* the classical force.

**Recapitulation**

This essay has attempted to address, from a modern perspective, the motion of a particle. Quantum mechanically, motion consists of a series of localizations due to repeated interactions that, taken close to the limit of the continuum, yields a world-line. If a force acts on the particle, its probability distribution is accordingly modified. This must also be true for macroscopic objects, although now the description is far more complicated by the structure of matter and associated surface physics. The motion of macroscopic objects, as was illustrated in the context of electrostatic forces, is governed by the quantum mechanics of its constituent particles and their interactions with each other. The result may be characterized as: collective quantum mechanical motion *is* classical motion.

Since electromagnetic forces may be represented as a gauge field, electrostatic forces arise from the non-constant phase character of the electric field affecting many-particle wave functions. The example used was that of the force on a charged or uncharged metallic, conducting sphere placed in an initially constant and uniform electric field.

As was written in the introduction, there is little that is new in this essay. On the other hand, quantum mechanics is widely viewed as being imposed on the well-understood classical world of Newtonian mechanics and Maxwell's electromagnetism. This dichotomy is part of the pedagogy of physics and leads to much cognitive dissonance. In the end, there is no classical world; only a many-particle quantum mechanical one that, because of localizations due to environmental interactions, allows the emergence of the classical world of human perception. Newtonian mechanics and Maxwell's electromagnetism should be viewed as effective field theories for the "classical" world.

# References and Notes

[1] Lawrence Sklar, *Space, Time, and Spacetime* (University of California Press, Berkeley 1974), p. 328.

[2] R. P. Feynman, *Phys. Rev.* **76**, 749 (1949).

[3] This is best exemplified by the path integral formulation of non-relativistic quantum mechanics. The latter also has the virtue of explicitly displaying the non-local character of quantum mechanics.

[4] The use of the term "localization" is deliberate. There is no need to bring in the concept of measurements with its implicit assumption of the existence of an "observer." It is not necessary that the fact that an interaction has occurred somehow enter human consciousness in order for the particle to be localized in space and time. The argument that it must enter human consciousness has been used, for example, by Kemble [E. C. Kemble, *The Fundamental Principles of Quantum Mechanics with Elementary Applications*, Dover Publications, Inc., 1958, p. 331] who states that "If the packet is to be reduced, the interaction must have produced knowledge in the brain of the observer. If the observer forgets the result of his observation, or loses his notebook, the packet is not reduced." It is not our purpose here to enter into a discussion of quantum measurement theory, but interpretations such as that expressed by Kemble often—but not always—rest on a lack of clarity as to what the wave function is assumed to represent. That is, whether the wave function applies to a single system or only to an ensemble of systems. While the ensemble interpretation has proven conceptually quite valuable in a number of expositions of measurement theory, it is difficult to understand how the wave function cannot apply to an individual system given the existence of many interference experiments using a series of *individual* electrons—where each electron participating in the production of the interference pattern must interfere with itself. Perhaps the most well known attempt to bring consciousness into quantum mechanics is that of Eugene Wigner. The interested reader is referred to Wigner's book *Symmetries and Reflections* (Indiana University Press, Bloomington & London 1967), Section III and references therein.

In many circumstances a "measurement" done by an "observer" can be replaced by the role of the environment. The effect of such environmental decoherence has been discussed by Zurek and Halliwell: W. H. Zurek, "Decoherence and the Transition from Quantum to Classical", *Physics Today* (October 1991); J. J. Halliwell, "How the quantum universe became classical", *Contemporary Physics* **46**, 93 (March-April 2005).

[5] Because the discussion to follow will give a different picture of a particle path, this is a good point to illustrate how motion is often described in quantum mechanics. Bohm [D. Bohm, *Quantum Theory*, Prentice-Hall, Inc., N.J. 1961, p.137.] in describing how a particle path is produced in a cloud chamber maintains that ". . . when the electron wave packet enters the chamber, it is quickly broken up into independent packets with no definite phase relation between them . . . the electron exists in only *one* of these packets, and the wave function represents only the *probability* that any given packet is the correct one. Each of these packets can then serve as a possible starting point for a new trajectory, but each of these starting points must be considered as a separate and distinct possibility, which, if realized, excludes all others." If the particle has large momentum, ". . . the uncertainty in momentum introduced as a result of the interaction with the atom results in only a small deflection, so that the noninterfering packets all travel with almost the same speed and direction as that of the incident particle." [emphasis in the original]

[6] There is a considerable—and quite interesting—literature dealing with repeated "measurements" of a particle and what is known as "Turing's Paradox" or the "Quantum Zeno Effect." See, for example: B. Misra and E. C. G. Sudarshan, *J. Math. Phys.* **18**, 756 (1977); D. Home and M. A. B. Whitaker, *Ann. of Phys.* **258**, 237 (1997), lanl.arXiv.org, quant-ph/0401164..

[7] The term "visible" is put in quotes as a short-hand for the physical processes involved: the interaction of the particle needed to localize it on the space-like hypersurface and the detection of that interaction by the observer. It should also be emphasized that localization is in both space and *time*. Just as localization in space to dimensions comparable to the Compton wavelength corresponds to an uncertainty in momentum of ~$mc$, localization in time must be $\geq h/mc^2$ if the uncertainty in energy is to be less than or equal to the rest mass energy. For electrons this corresponds to $\geq 10^{-20}$ second.

[8] Rudolph Haag, *Quantum Theory and the Division of the World*, *Mind and Matter* **2**, 53 (2004).

[9] If one uses only positive energy solutions of the Dirac equation to form a wave packet, the probability of finding a particle outside the light cone nowhere vanishes, although the propagator becomes very small for distances greater than the Compton wavelength $\hbar/mc$.

[10] L. O'Raifeartaigh, *The Dawning of Gauge Theory* (Princeton University Press, Princeton, New Jersey 1997), pp. 24-37.

[11] G. E. Marsh, "Charge, Geometry, and Effective Mass", *Found. Phys.* **38**, 293-300 (2008).

[12] I. J. R. Aitchison and A. J. G. Hey, *Gauge Theories in Particle Physics* (Institute of Physics Publishing, Bristol and Philadelphia 1993), 2nd edition.

[13] R. W. Robinett, "Quantum mechanical time-development operator for the uniformly accelerated particles", *Am. J. Phys.* **64**, 803-808 (1996).

[14] G. Vandegrift, "Accelerating wave packet solution to Schrödinger's equation", *Am. J. Phys.* **68**, 576-577 (2000).

[15] J. R. Smith, "Theory of Electronic Properties of Surfaces", contained in: R. Gomer, ed., *Interactions on Metal Surfaces* (Springer-Verlag New York 1975), The term "jellium" was apparently first introduced by Conyers Herring.

[16] N. D. Lang and W. Kohn, "Theory of Metal Surfaces: Work Function", *Phys Rev.* B**3**, 1215-1223 (1971).

[17] R. G. Forbes, "Charged Surfaces, Field Adsorption, and Appearance-Energies: an Unsolved Challenge", *Journal de Physique* IV (Colloque C5, supplement au Journal de Physique III, Vol. 6 septembre), C-25—C-30 (1996).

[18] E. Hult, et al., "Density-functional calculation of van der Waals forces for free-electron-like surfaces", *Phys. Rev.* B **64**, 195414.